\documentclass[11pt]{article}
\usepackage{moriond,epsfig}

\def\lsim{\raise0.3ex\hbox{$<$\kern-0.75em\raise-1.1ex\hbox{$\sim$}}}
\def\gsim{\raise0.3ex\hbox{$>$\kern-0.75em\raise-1.1ex\hbox{$\sim$}}}

\def\be{\begin{equation}}
\def\ee{\end{equation}}
\def\bea{\begin{eqnarray}}
\def\eea{\end{eqnarray}}

\begin{document}
\vspace*{4cm}

\title{JETS IN HEAVY ION COLLISIONS.}

\author{
 C.A. Salgado
}

\address{
CERN, Theory Division, CH-1211 Geneva, Switzerland
}

\maketitle\abstracts{
High energetic particles traversing a dense medium lose a sizable 
part of their energy in form of gluon radiation. As a result, the rate of
high-$p_t$ particles is expected to be suppressed in heavy ion collisions
with respect to the proton case. Recent experimental data from RHIC
strongly support this scenario. 
This 
allows to study the properties of the medium
by the amount of {\it jet quenching} it produces. The angular
dependence of the radiation is modified in the medium in a characteristic
way. This provides another tool to study the medium properties
in a more differential measurement.
}

The evolution, in the vacuum, of a quark or gluon produced 
with high virtuality is given by gluon radiation and described by
DGLAP-like evolution equations. In this way,
the study of jets (production, shapes, etc...) in proton--proton, lepton--proton
or $e^+e^-$ collisions, give very valuable
information about the dynamics of QCD. In heavy ion collisions,
where a dense state is expected to be formed, this high energetic quark or
gluon would traverse  the produced medium before hadronization. In this case, 
the interaction with the medium
modifies the gluon radiation, and hence the evolution. 
This has been proposed \cite{quench} long ago as a tool to study the properties
of the produced medium. One of the main predictions is the suppression of
particles produced at high-$p_t$ in heavy ion collisions. Recent experimental 
data from RHIC found a strong suppression of the particle spectrum at 
large-$p_t$ in central AuAu collisions \cite{rhicAA}. 
This same spectrum is not suppressed
but enhanced in dAu collisions, also measured in RHIC \cite{rhicdA}. 
The conclusion is that
the suppression in AuAu is not due to initial state effects but to strong
final state interactions.

The modified gluon
radiation due to multiple (coherent) scatterings has been computed
by several groups, in the multiple soft scattering limit \cite{bdmps},
in the hard scattering approximation \cite{glv}, they both are 
limiting cases of a general path integral formalism \cite{urs}. 
(For an alternative formalism, based on a twist--expansion see 
Ref. \cite{xinnian}).
Once the
medium-modified gluon radiation is known, the modified evolution can, in 
principle, be computed. 
In Fig. 1, the $k_\perp$--integrated 
spectrum of gluons radiated by a high energetic quark
is plotted for different media in the multiple soft scattering approximation 
\cite{noso2} as a function of the energy of the emitted gluon $\omega$. 
The quantities which characterize the medium are
the length $L$ and the transport coefficient $\hat q$. This last quantity
is given by the amount of transverse momentum that the parton gets
per unit lento when suffering multiple scattering with the medium \cite{noso2}.
The spectrum can be written as a function of only two variables,
$\omega_c=\hat q L^2/2$ and $R=\omega_c L$.

\begin{figure}[tb]
\centering{\epsfxsize=9.5cm\epsfbox{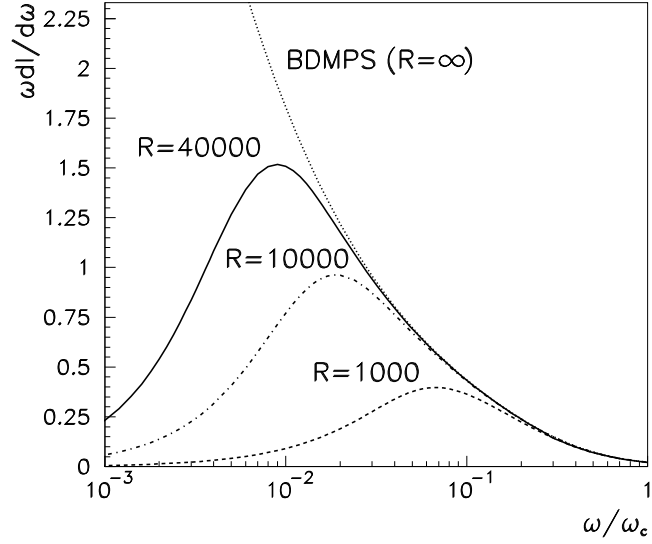}}
\vspace{0cm}
\caption[a]{Medium-induced gluon energy distribution $\omega dI/d\omega$ 
for different media \cite{noso2}.
The result for $R\to\infty$ coincides 
with the BDMPS \cite{bdmps} case.} 
\end{figure}

The first observation is that the energy distributions from Fig. 1 
are cut-off at small values of
the energy of the emitted gluon $\omega$. The reason is the following, the
typical transverse 
momentum of gluons emitted coherently is $k_\perp^2\sim \hat qL$. 
The kinematical constraint for the $k_\perp$--integration,
$k_\perp\leq \omega$ is, then,  relevant for
values of $\omega/\omega_c\lsim\sqrt{2/R}$. The positions of the maxima
in the figure agrees with this estimate. One sees also that  the BDMPS
case \cite{bdmps} is recovered \cite{noso2} for $R\to\infty$. This corresponds
to remove the upper limit in  the $k_\perp$-integration.

The spectrum of gluons radiated outside a cone of angle $\Theta$ is given
by \cite{urs,noso2}
$dI/d\omega-dI^{R'}/d\omega$, where $R'=R\cos \Theta$. This
allows to compute, for a medium, the additional energy radiated 
outside a cone,
$\Delta E(\Theta)
=\int d\omega \ \omega (dI/d\omega-dI^{R'}/d\omega)$. 
In Fig. 2 this energy is plotted
as a function of the angle. The induced gluon
emission has a characteristic angular structure. This structure translates
into modification of jet shapes due to the medium, the so
called {\it jet broadening} 

\begin{figure}[tb]
\centering{\epsfxsize=8cm\epsfbox{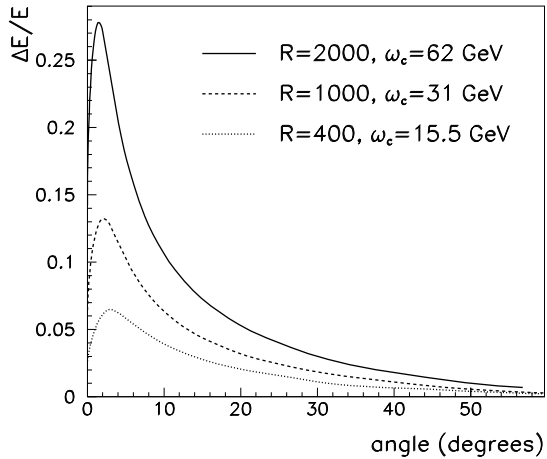}}
\vspace{0cm}
\caption[a]{The average energy loss radiated outside an
angle $\Theta$ for a quark jet of energy E=100 GeV.}
\end{figure}

One of the most important experimental results from RHIC is the suppression
of high-$p_t$ particles in nucleus-nucleus collisions \cite{rhicAA}. The recent
experimental data on dAu collisions \cite{rhicdA}, found
an enhancement  of particle production at $2\lsim
p_t\lsim 10$. This  indicates that the suppression in AA collisions
is not of initial state origin. The final state high-$p_t$ suppression due
to energy loss of the partons traveling through the produced medium is 
the most natural explanation for this suppression.

An simple way of computing the effect of energy loss in heavy ion collisions is
through the so called quenching factor \cite{bdms}

\begin{equation}
 Q(p_t)=
{{d\sigma^{MED}(p_t)/ dp^2_t}\over
{d\sigma^{VAC}(p_t)/ dp^2_t}}=
\int d\epsilon P_E(\epsilon)\left(
{d\sigma^{VAC}(p_t(1-\epsilon))/ dp^2_t}\over
{d\sigma^{VAC}(p_t)/ dp^2_t}\right)\ ,
\label{eqqf}
\end{equation}

\noindent
where the medium spectrum is computed by a shift, $\epsilon p_t$, in the 
vacuum spectrum, $d\sigma^{VAC}$, and weighted by 
the probability, $P_E(\epsilon)$, that the hard parton loses a fraction
$\epsilon$ of its initial energy $E$. The {\it quenching weight}, 
$P_E(\epsilon)$, is usually computed in the
independent gluon emission approximation \cite{bdms}

\begin{eqnarray}
  P_E(\epsilon) = \sum_{n=0}^\infty \frac{1}{n!}
  \left[ \prod_{i=1}^n \int d\omega_i \frac{dI(\omega_i)}{d\omega}
    \right]
    \delta\left(\epsilon -\sum_{i=1}^n {\omega_i \over E} \right)
    \exp\left[ - \int d\omega \frac{dI}{d\omega}\right]\, .
\label{eqpe}
\end{eqnarray}

Computing the suppression  by means of
medium-modified fragmentation functions gives very similar results
\cite{noso2}, so, expression (\ref{eqqf}) gives a good estimation of the
effect.
In Fig. 3 we compare the quenching factors that we obtain for $\omega_c$=25 GeV
and $R$=1000 with the experimental data from PHENIX Collaboration \cite{rhicAA}
both for quark and gluon jets. For the vacuum spectrum we use the fit
provided by PHENIX,  

\begin{equation}
\frac{d\sigma^{VAC}}{dp_t^2}=
\mbox{const}\cdot \left(1.71 + p_t [\mbox{GeV}]\right)^{-12.44} .
\end{equation}

\begin{figure}[tb]
\centering{\epsfxsize=10cm\epsfbox{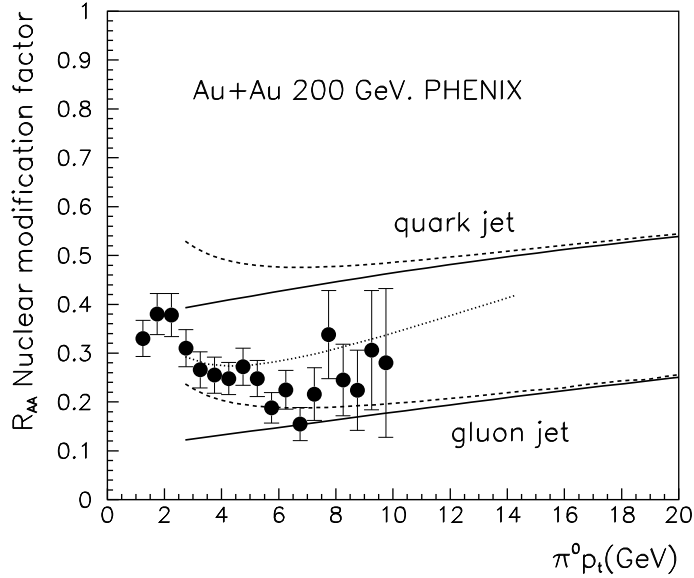}}
\vspace{0cm}
\caption[a]{Quenching factors for quarks and gluons compared with PHENIX
data \cite{rhicAA}
on $\pi^0$ suppression in AuAu collisions. Solid lines are computed using
the spectrum of Fig. 1 for $R$=1000, $\omega_c$=25 GeV. Dashed lines take into
account finite energy cuts. Dotted line takes into account the $p_t$-dependence 
of the fractional contribution from quarks and gluons to the final pions.}
\end{figure}

The spectrum of Fig. 1 is computed in the approximation $\omega\ll E$, $E$ being
the energy of the hard parton. This is a
common feature with other multiple scattering approaches \cite{bdmps,glv}.
In the kinematical range covered by RHIC experiments, $p_t\lsim 10$ GeV, 
finite energy effects could appear. In order to check the sensitivity of our 
results to the region $x\equiv \omega/E\sim 1$ we impose the kinematical cut
for the quenching weights $P_E(x)=0$ for $x> 1$ \footnote{The finite energy
corrections to the emission vertex are the same as in the vacuum
\cite{bdmsfin}. So, these finite energy corrections could be taken into account
by simply multiplying the spectrum by the corresponding splitting function for
quarks or gluons. In the case of the quarks, we have checked that the results
are similar to the ones we obtain here. However, in the
case of gluon jets, the configuration is symmetric and
the probability of large longitudinal 
momentum transfer is large for $x\to 1$. Including this effect is not
simple in the present formalism, as the hadronization
of emitted gluons is not included.}.
The energy of the initial parton is not
equal to $p_t$; in our estimation, we assume that 
a pion takes in average a fraction $\sim$0.7 
($\sim$0.55) of the initial quark (gluon) energy \cite{kariheli}.   
The effect of these finite energy cuts is plotted in Fig. 3. One can see
that these kinematical
cuts
have some importance for the smallest values of
$p_t$, however, they almost disappear for
$p_t\gsim$ 10 GeV. This conclusion depends
on the actual value of $\omega_c$; the larger the value is, the larger the 
finite energy effects are too.
An important observation, however,
is that finite energy corrections reduce the degree of jet quenching for
the small values of $p_t$. This changes the curvature of the suppression in 
agreement with the trend of experimental data. The contribution of quarks and 
gluons to the observed pion yield depends on $p_t$. For comparison, we have
computed the suppression taking into account the fractional 
contribution of quarks and gluons to the final particle from Ref. \cite{ramona}.
This gives us an estimation of the net effect and the slopes. A more refined
analysis, including nuclear parton distribution functions, fragmentation
function, etc.. is, however, needed  for a quantitative description.

\end{document}